\pdfoutput=1
\documentclass[a4paper]{aa}
\usepackage{graphicx,float}
\usepackage{latexsym,amsmath,amssymb}
\usepackage{natbib}
\usepackage[scriptsize,tight]{subfigure}
\bibpunct{(}{)}{;}{a}{}{,}
\usepackage[english]{babel}

\def \isdc        {INTEGRAL Science Data Centre, Universit\'e de Gen\`eve, Chemin d'Ecogia 16, CH-1290
                  Versoix, Switzerland}

\def \obsgen      {Observatoire de Gen\`eve, Universit\'e de Gen\`eve, Chemin des Maillettes 51, CH-1290
                Sauverny, Switzerland}

\def\dpnc      {D\'epartement de Physique Nucl\'eaire et Corpusculaire, Universit\'e de Gen\`eve, Quai Ernest-Ansermet 24, CH-1211 Gen\`eve 4, Switzerland}

\def \saclay {AIM Paris Saclay, CEA/DSM-CNRS/INSU-Universit\'e Paris Diderot, IRFU/Service d'Astrophysique, 91191 Gif-sur-Yvette, France}

\def \hcm {\hbox {\ifmmode $ atom cm$^{-2}\else atom cm$^{-2}$\fi}}
\def\approxgt{\mathrel{\hbox{\rlap{\lower.55ex \hbox {$\sim$}}
        \kern-.3em \raise.4ex \hbox{$>$}}}}
\def\approxlt{\mathrel{\hbox{\rlap{\lower.55ex \hbox {$\sim$}}
        \kern-.3em \raise.4ex \hbox{$<$}}}}

\newcommand {\degree} {$^{\circ}$}
\def\arcmin{\hbox{$^\prime$}}
\def\arcsec{\hbox{$^{\prime\prime}$}}

\newcommand{\hessj}{HESS~J1632-478}
\newcommand{\xmmu}{XMMU~J163219.9-474731}

\begin{document}

\title{HESS~J1632-478: an energetic relic}

\author{M. Balbo\inst{1, 2}, P. Saouter\inst{3}, R. Walter\inst{1, 2}, L. Pavan \inst{1, 2}, A. Tramacere\inst{1,2}, M. Pohl \inst{3}, J.-A. Zurita-Heras\inst{4}}

\offprints{Matteo.Balbo@unige.ch}

\institute{\isdc \and \obsgen \and \dpnc \and \saclay}

\date{Received 11 April  2010; Accepted 4 July 2010}

\authorrunning{M. Balbo, P. Saouter, R. Walter, et al.}

\titlerunning{HESS~J1632-478: an energetic relic}

\abstract
{}
{\hessj\ is an extended and still unidentified TeV source in the galactic plane.}
{In order to identify the source of the very high energy emission and to constrain its spectral energy distribution, we used a deep observation of the field obtained with XMM-Newton together with data from Molonglo, Spitzer and Fermi to detect counterparts at other wavelengths.}
{The flux density emitted by \hessj\ peaks at very high energies and is more than 20 times weaker at all other wavelengths probed. The source spectrum features two large prominent bumps with the synchrotron emission peaking in the ultraviolet and the external inverse Compton emission peaking in the TeV. \hessj\ is an energetic pulsar wind nebula with an age of the order of $10^4$ years. Its bolometric (mostly GeV-TeV) luminosity reaches 10\% of the current pulsar spin down power. The synchrotron nebula has a size of 1 pc and contains an unresolved point-like X-ray source, probably the pulsar with its wind termination shock. }
{}


\maketitle

\bigskip
\section{Introduction}

\hessj\ has been discovered as a diffuse very high energy (VHE) $\gamma$-ray source 
during the 2004-2006 survey of the inner Galaxy  \citep{2006ApJ...636..777A} performed
with the High Energy Stereoscopic System Cherenkov Telescope Array \citep[H.E.S.S.;][]{2004NewAR..48..331H}.

Though many galactic VHE sources are recognized to be supernova remnants, 
pulsars and pulsar wind nebulae, a small number of them still lack a clear 
identification. \hessj\ is one of them, even if tentatives to explain its nature 
have already been presented \citep{2006ApJ...636..777A, 2007Ap&SS.309....5W}.

A detailed VHE spectral and positional analysis of \object{HESS J1632-478}
has been reported in \cite{2006ApJ...636..777A}, based on 4.5 hours of H.E.S.S. observations.
The source best fit position is RA=16:32:09 DEC=--47:49:12 (J2000), placing it in the direction 
of the near 3 kpc arm tangent in the galactic plane.

The TeV source is extended with a semi-major axis of  $(12\pm 3)\arcmin$,
forming an angle of $21\pm 13^\circ$ with respect to the positive galactic longitude axis.
The VHE spectrum, between 0.2 and 4.5 TeV, can be fitted with a powerlaw model, yielding
a photon index $\Gamma=2.12 \pm 0.20$ and a flux above 200~GeV of
 $(28.7 \pm 5.3) \times 10^{-12}$ ph\ cm$^{-2}$ s$^{-1}$.

To understand \object{HESS J1632-478}, we collected multi-wavelength data from 
Fermi, XMM-Newton, Spitzer and Molonglo, to construct its spectral energy distribution
and discuss the emission mechanisms.

\section{X-ray Counterpart}\label{secother}

\subsection{XMM-Newton observation and data analysis}\label{sect:xmmobs}

The XMM-Newton observatory \citep{jansen01aa} includes three
1500~cm$^2$ X-ray telescopes each with a European Photon Imaging
Camera (EPIC) at the focus.  Two of the EPIC imaging spectrometers
use MOS CCDs \citep{turner01aa} and one uses pn CCDs
\citep{struder01aa}. 

In the period from August to September 2008, XMM-Newton performed 9 
observations of a field centered close to \hessj, collecting data for a total
of 92~ks. The EPIC observations, listed in table \ref{ObservationID},
used the thin optical blocking filter. The EPIC pn camera was operated 
in large-window mode.

Two other XMM-Newton observations of the same field were collected in 2003 and
2004. As these observations used the medium optical blocking filter and did not use 
the central MOS CCD and the 2004 one was heavily affected by an outburst of 4U 1630-472, 
we did not combine these observations with those of 2008.

\begin{table}[h] 
\caption{Observation identifiers and date for all XMM-Newton observations of 2008 including the source  \hessj. The actual 
and filtered exposure times in seconds are given for the two EPIC MOS and for the pn camera respectively.}
\label{ObservationID}
\begin{tabular}{c |c |r | r| r| r}
\hline \hline \noalign{\smallskip} 
Obs. ID & Date & \multicolumn{1}{c|}{MOS} & \multicolumn{1}{c|}{ MOS} & \multicolumn{1}{c|}{pn} & \multicolumn{1}{c}{pn}\\
 & & \multicolumn{1}{c|}{act. s} & \multicolumn{1}{c|}{filt. s} & \multicolumn{1}{c|}{act. s} & \multicolumn{1}{c}{filt. s} \\
\noalign{\smallskip\hrule\smallskip}
556140101 & 14.08.08 &  10621& 10621& 9834& 5800\\
556140201 & 16.08.08 &  9620& 7500& 8034& 5900\\
556140301 & 18.08.08 &  8619& 7500& 7034& 4500\\
556140401 & 20.08.08 &  12829& 12821& 11234 & 11000\\
556140501 & 21.08.08 &  4951& 4951& 2239& 1900\\
556140601 & 22.08.08 &  14027& 14027 & 12398& 12398\\
556140701 & 24.08.08 &  10621& 10621& 9834& 6700\\
556140801 & 26.08.08 &  11219& 11219& 9634& 9500\\
556141001 & 17.09.08 &  9118& 8200& 7534 & 4500\\
\noalign{\smallskip\hrule\smallskip}
\end{tabular}
\end{table}

Standard data reduction procedures were applied to each observation using the 
XMM~\textsc{Science Analysis Software} (SAS Version 9.0.0)\footnote{http://xmm.esac.esa.int/sas/current/}. 
Source positions were derived using the SAS task \texttt{edetect\_chain} with 
the EPIC MOS and pn data. Only well calibrated single pixel events were selected for the pn 
CCD and single and double events for the MOS CCDs. Known hot, or flickering, pixels and 
electronic noise were rejected. Events were further screened with the conservative 
\textsc{flag=0} criteria and selected for the energy ranges ($0.2-12$ keV for 
MOS and $0.2-15$ keV for pn). Periods with enhanced background -- 
soft proton flares (EPIC pn count rate above 10 keV larger than 0.7 to 4.5 ct/s, depending on 
the observation) -- 
were disregarded in the analysis, resulting in total filtered exposures of 87 ks 
and 62 ks for the MOS and pn cameras respectively. 

Events were finally selected in the source and background regions. Circular regions have been chosen
for point sources. The regions used for the likely counterpart to \hessj\ are described in section \ref{sect:hess}.

\subsection{Serendipitous sources}

In the nine XMM-Newton observations of 2008, three point-like sources are always clearly detected: 
\object{IGR~J16320-4751}, \object{AX~J1632.8-4746} and \object{\xmmu}, by order of decreasing X-ray flux. The 
EPIC derived positions of these sources are listed in table \ref{tab:Sources}.

\begin{table}[t] 
\caption{Source identifier, name, positions and positional statistical error ($\Delta$) derived from XMM-Newton data. 
The systematic error on the XMM-Newton attitude reconstruction is of the order of 1.5\arcsec$^2$.}
\label{tab:Sources}
\begin{tabular}{l |l |c | c| c}
\hline \hline \noalign{\smallskip} 
ID & Name & RA & DEC &$\Delta$ \\
\noalign{\smallskip\hrule\smallskip}
A&{\tiny \xmmu}&16:32:19.9&-47:47:31& 0.19\arcsec\\
B&AX~J1632.8-4746&16:32:48.2& -47:45:04& 0.13\arcsec\\
C&IGR~J16320-4751&16:32:01.9&-47:52:29& 0.87\arcsec\\
Ee&\hessj&16:32:07.8&-47:49:09&1\arcsec\\
Ep&\hessj&16:32:08.8&-47:49:01&0.5\arcsec\\
\noalign{\smallskip\hrule\smallskip}
\end{tabular}
\end{table}
\footnotetext[2]{http://xmm.vilspa.esa.es/docs/documents/CAL-TN-0018.pdf}

The XMM-Newton data of the high-mass X-ray binary IGR~J16320-4751 will be presented elsewhere.
In these data, IGR~J16320-4751 features column densities in the range $(8 - 10)\times 10^{22}$ cm$^{-2}$.

AX J1632.8-4746 is detected by XMM-Newton as a constant source with $(14.5 \pm 0.3) \times 10^{-2}~\rm{counts~s}^{-1}.$
Its spectrum can be well reproduced with the 
absorbed emission of a hot diffuse gas \citep[\texttt{mekal,}][]{kaastra92} yielding 
N$_{\rm H}$ = $(6.2 \pm 0.2)\times 10^{22}\textrm{cm}^{-2}$, kT = $ 2.5 \pm 0.12$~keV 
and F$_{2-10 {\rm keV}}=(3.25\pm 0.1) \times 10^{-12}$ erg cm$^{-2}$ s$^{-1}$.
The ASCA data \citep{2001ApJS..134...77S} were described with an absorbed  power-law 
resulting in a folded model comparable to that derived with XMM-Newton.

The light-curve of \xmmu\ does not show any variation. Its spectrum is 
also well reproduced with an absorbed \texttt{mekal} model with N$_{\rm H} =(5.5 \pm 0.3)\times 10^{22}$ cm$^{-2}$, 
kT $= 2.2 \pm 0.1$~keV and F$_{2-10 {\rm keV}}= (1.1\pm 0.1) \times 10^{-12}$ erg cm$^{-2}$ s$^{-1}$.

The spectra of both AX J1632.8-4746 and \xmmu\ are consistent with these of
massive stars. If these sources are at a distance $\gtrsim 2$ kpc, their X-ray 
luminosity ($\lesssim 10^{33}$ erg/s) is consistent with the expectations \citep[$10^{30-33}$ erg/s,][]{Cassinelli1981}. 
We find possible infrared point-like counterparts to these two XMM sources in the Spitzer surveys \citep{2003PASP..115..953B, Carey-MIPSGAL} 
at distances of 3.3\arcsec and 2.0\arcsec respectively.

\subsection{\hessj\ Counterpart\label{sect:hess}}

None of the point-like X-ray sources discussed above is a likely counterpart of the extended source detected 
in the HESS survey, therefore we merged together all the observations to obtain a better sensitivity.
To construct mosaic images of the MOS and pn data, described in Sect. \ref{sect:xmmobs}, we 
used the \texttt{emosaic} command. The mosaic of the count images has been divided by the mosaic 
of the individual exposure maps calculated with the \texttt{eexmap} command.

\begin{figure}[h!] 
\begin{center}
\includegraphics[width=6cm]{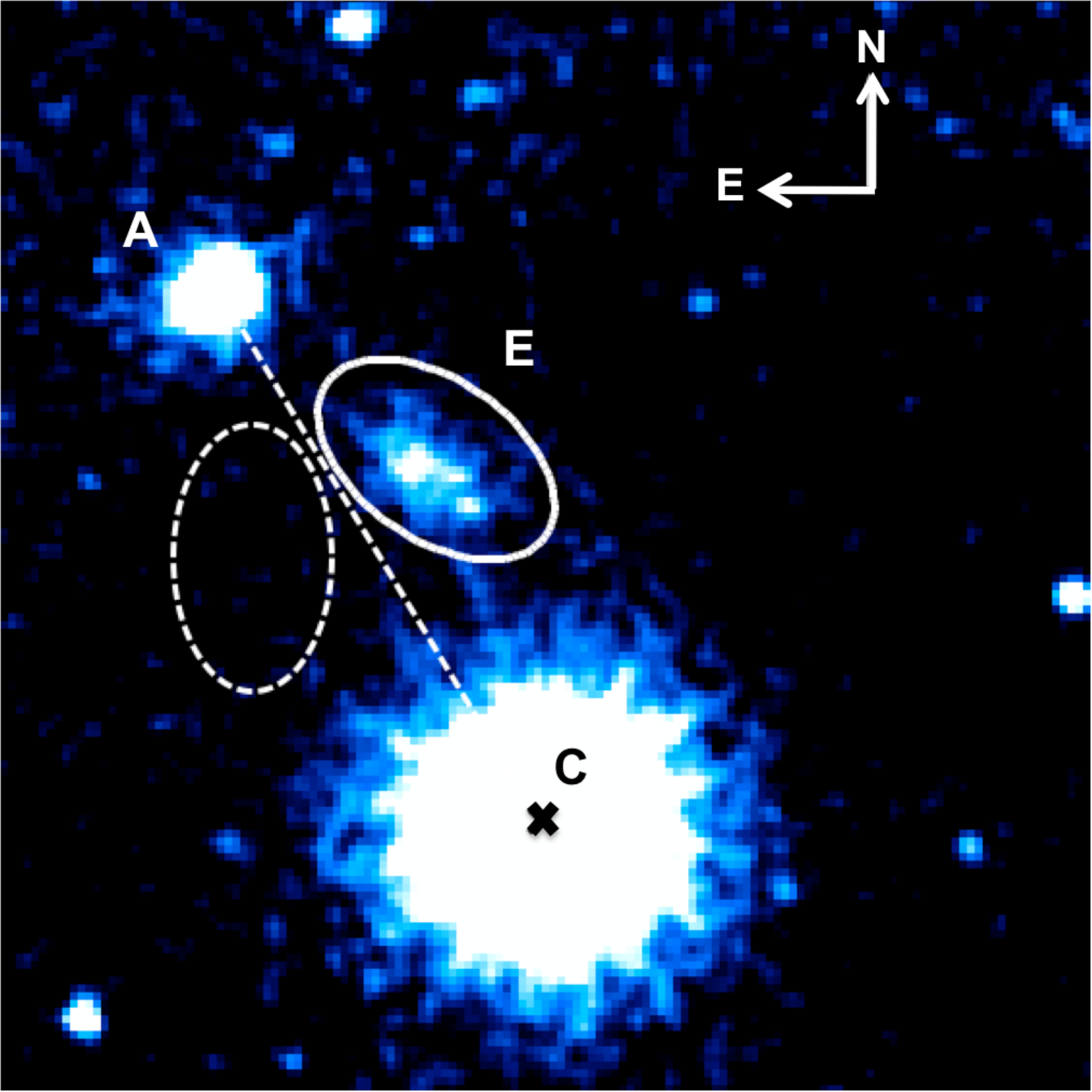}
\includegraphics[width=6cm]{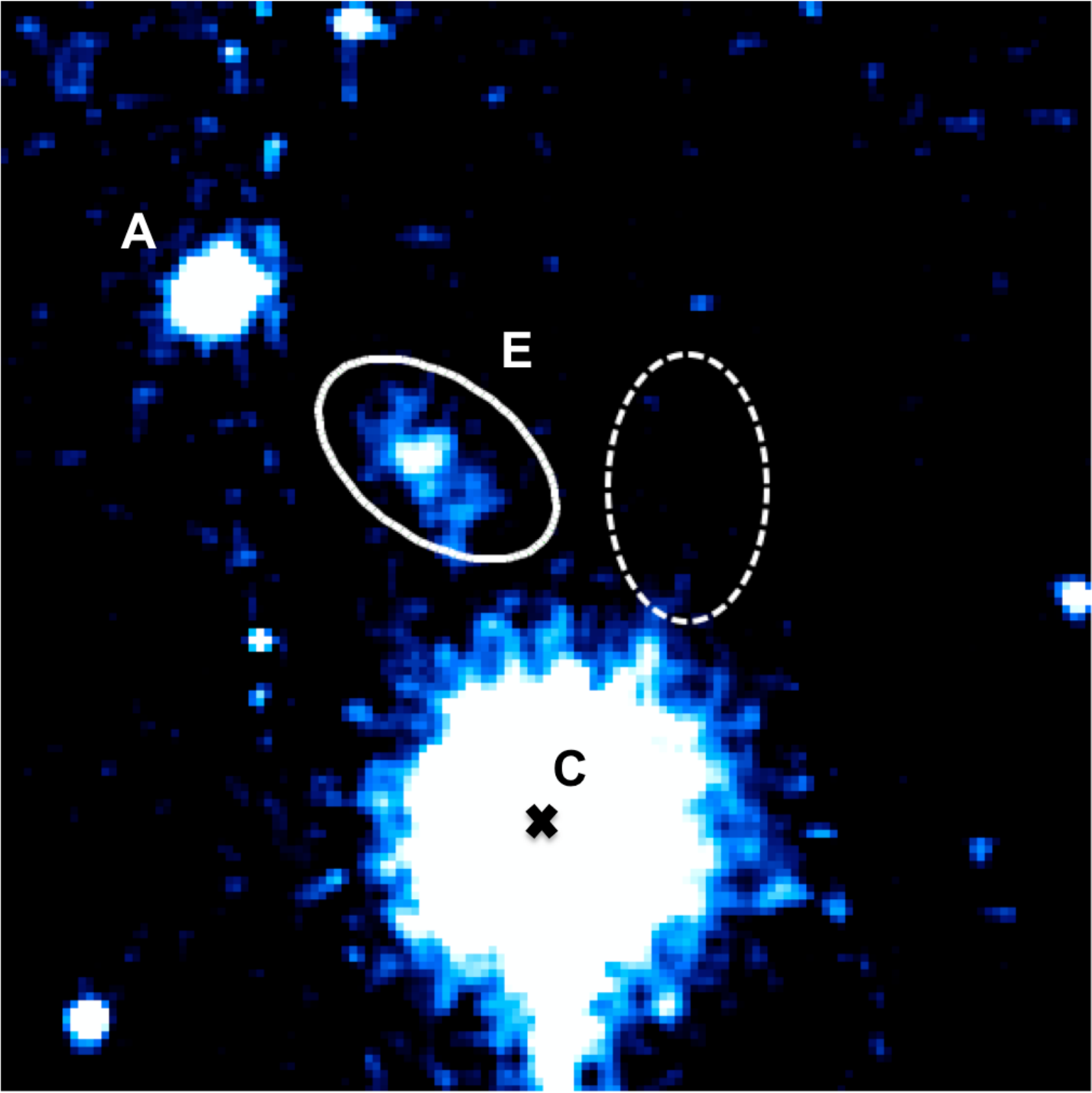}
\includegraphics[width=6cm]{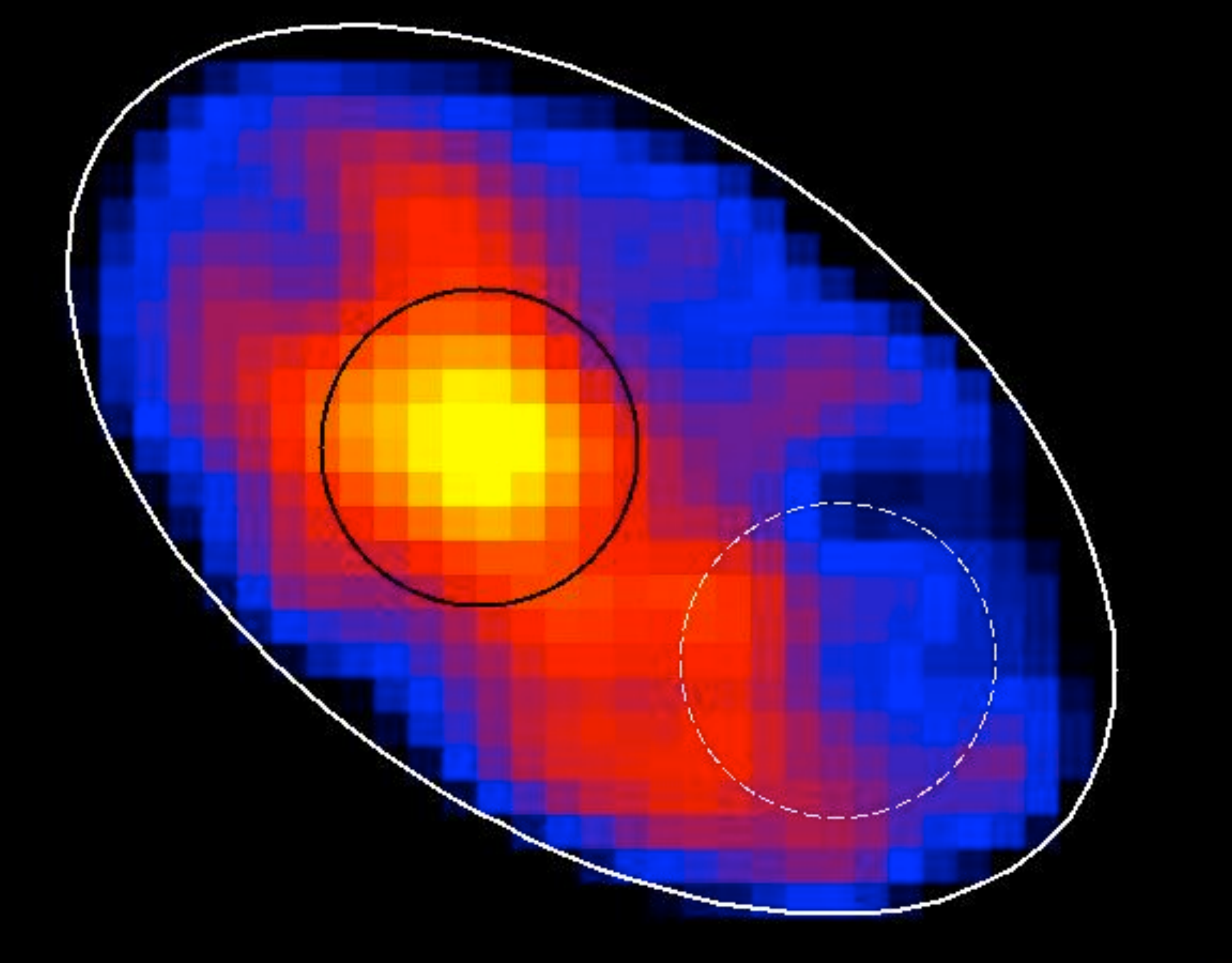}
\end{center}
\caption{The upper and middle panels show XMM-Newton EPIC mosaic images in the energy band 0.2-12 keV for the 
MOS and 0.2-15 keV for pn CCDs, respectively. A smoothing algorithm was applied. The height of these images is 10 arcmin. Ellipses 
indicate the source (continuous) and background (dashed) extraction regions for the extended source. The enlarged bottom image extracted from the MOS1 camera, shows the extraction regions
for the point-like (black) and extended (white) sources respectively. Events in the black circle are not included in the spectrum 
of the extended source. The dashed white circle shows the region used to select the ``background'' events of the point-like source.}
\label{fig:MOSPNExtraction}
\end{figure}

The resulting MOS and PN mosaic images are presented in Fig. \ref{fig:MOSPNExtraction}. An extended source 
is detected in the two mosaic images (source E). The source profile features an unresolved point-like component 
(source Ep in table \ref{tab:Sources}), with a width $(\sigma=6.8\pm 0.6)\arcsec$ consistent with the XMM-Newton 
point spread function, in addition to an extended component (source Ee in table \ref{tab:Sources}), close to the detection 
limit. The point and extended sources have significances of 15 and 18 $\sigma$ respectively. The extended source could be 
represented by a two dimensional gaussian profile with $1\sigma$ semi-axes of $(32.0\pm1.2)\arcsec$ and $(15.3\pm0.7)\arcsec$. 
The angle of the major axis of the extended source is $(55\pm2)\degr$ counted anti-clockwise from the north direction. 
 
To extract the X-ray spectra of the point-like and extended source, we first defined the source (Fig. \ref{fig:MOSPNExtraction} bottom) and background (Fig. \ref{fig:MOSPNExtraction} top and middle)
extraction regions in the mosaic images and selected the events in these regions for each individual observation. 
The elliptical region used to select the extended source excludes a circle around the point-like  source (Fig. \ref{fig:MOSPNExtraction} bottom).
The ancillary spectral response file for the extended source was generated with the \texttt{extendedsource=YES} option of the \texttt{arfgen} tool.

For the MOS CCDs, the background extraction region was chosen in order to minimize the possible contributions 
from IGR~J16320-4751 and \xmmu\ (Fig. \ref{fig:MOSPNExtraction} top). A different background extraction region 
was chosen for the PN camera (Fig. \ref{fig:MOSPNExtraction} middle) to avoid falling into the gap between the CCD 
chips when extracting the background in individual observations. 

We applied these extraction regions to each observation separately. Source and background spectra and response matrices 
were extracted for each observation. As the source is at the same position in the detector plane for all the 2008 observations, 
the resulting spectra were combined together using \texttt{addspec} (without error propagation, i.e. using Poisson statistics) to 
obtain a single merged spectrum for each camera. 

\begin{table}[h]
\caption{Best fit parameters for an absorbed power-law model fitted simultaneously to the three EPIC spectra on the
point-like and extended sources. Reported errors correspond to $\mathbf{1\sigma}$ uncertainties (68\%). The 2-10 keV fluxes are corrected for the absorption.}
\label{tab:fit1}
\begin{tabular}{c|c|c|l}
\hline \hline \noalign{\smallskip} 
    Parameter   &    Point &Extended& Unit\\
\noalign{\smallskip\hrule\smallskip}
    N$_H$                         &$13_{-4}^{+6}$            &$11_{-2.7}^{+2.2}$   &$10^{22}$ cm$^{-2}$\\
    $\Gamma $                 &$2.6_{-0.8}^{+1.3} $   &$3.4_{-0.8}^{+0.6} $   &\\
    F$_{\rm 2-10 keV}$   &$2.3_{-1.0}^{+0.3} $   &$4.3 _{-0.4}^{+0.8} $  &$10^{-13}$ erg cm$^{-2}$ s$^{-1}$\\
    $\chi^2_\nu$               &1.8                                 &1.9              &\\
\noalign{\smallskip\hrule\smallskip}
\end{tabular}
\end{table}

\begin{figure}[h]
\centering
\includegraphics[angle=270, width=\linewidth]{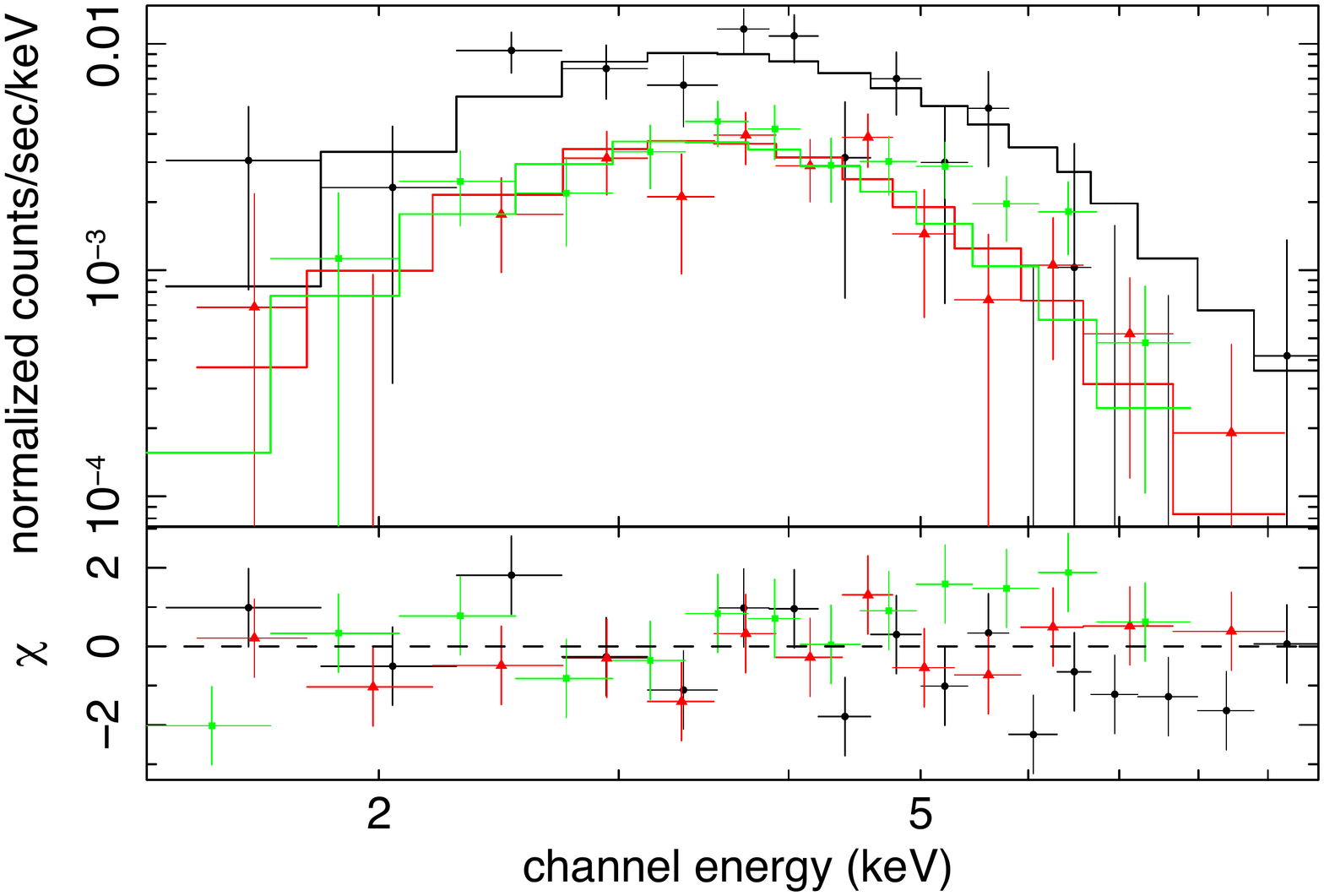}
\vspace{-10mm}

\includegraphics[angle=270, width=\linewidth]{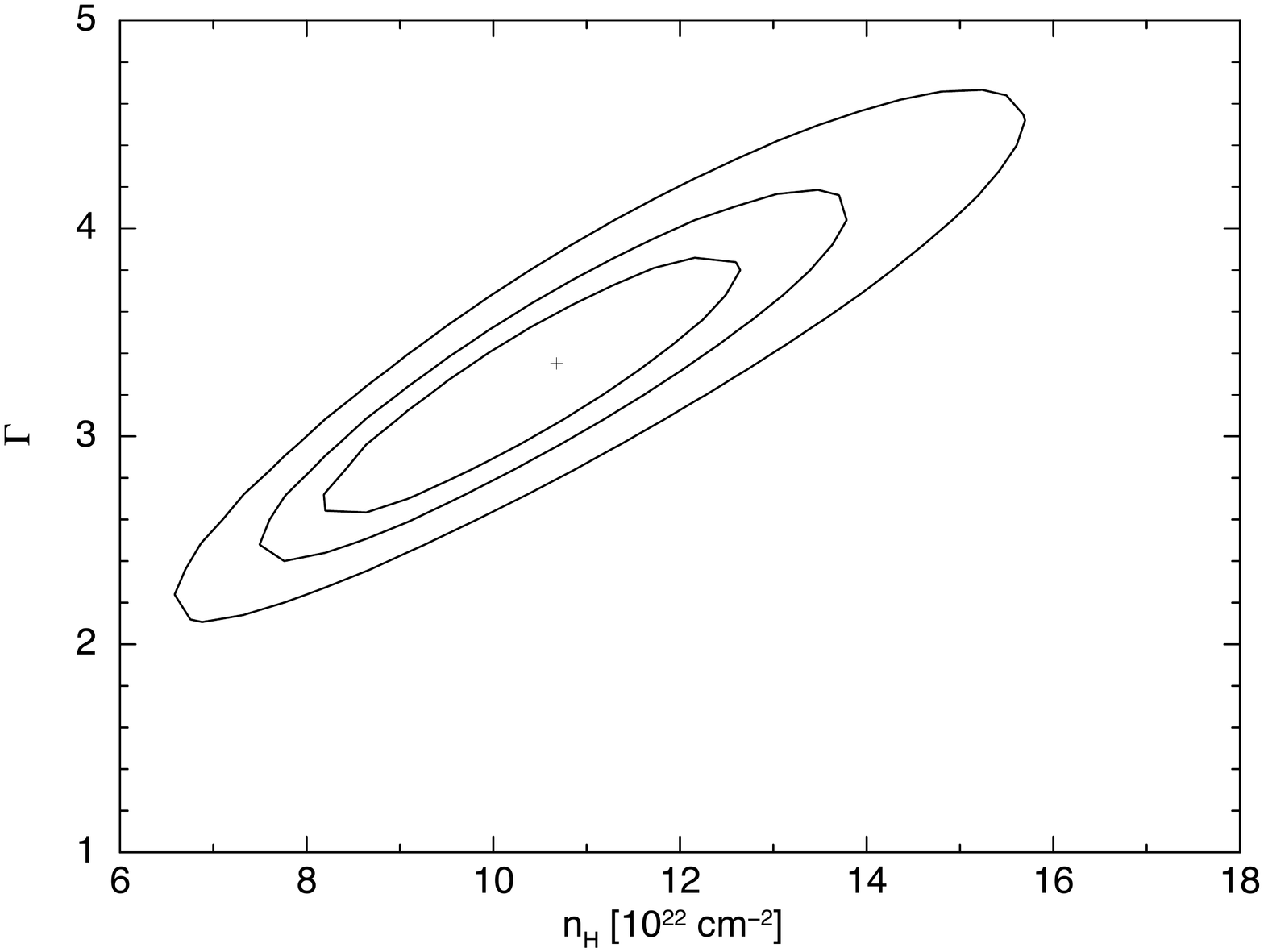}
\caption{Top: Best power-law fit of the spectra of the extended X-ray source obtained from the three EPIC cameras. Bottom: 68\%, 90\% and 99\% confidence contours for the N$_H - \Gamma$ parameters.}
\label{fig:bestfit}
\end{figure}

The resulting spectra have been analyzed with \texttt{XSPEC} (version 11)\footnote[3]{http://heasarc.nasa.gov/docs/xanadu/xspec/xspec11/}. We fitted 
simultaneously the merged EPIC spectra obtained for the three cameras with an absorbed power-law model (Fig. \ref{fig:bestfit}). The resulting $\chi^2_\nu$ 
(table~\ref{tab:fit1}) are large, probably because the count rate uncertainties related to the background subtraction are underestimated.  

The absorbed and unabsorbed fluxes of the extended source are F$_{2-10keV} = 1.4 \times 10^{-13}$ erg cm$^{-2}$ s$^{-1}$ and 
$4.3 \times 10^{-13}$ erg cm$^{-2}$ s$^{-1}$, respectively. 

We also attempted to fit the point-like source with an absorbed black-body model. The derived temperature and emitting radius are 1 keV and 22-43 m 
(for a distance of 3 kpc). These parameters are not meaningful as a power-law model fits the data as well and is probably a better representation of the 
emission at the termination shock.

\begin{figure}[t]
 \begin{center}
 \includegraphics[width=0.45\textwidth]{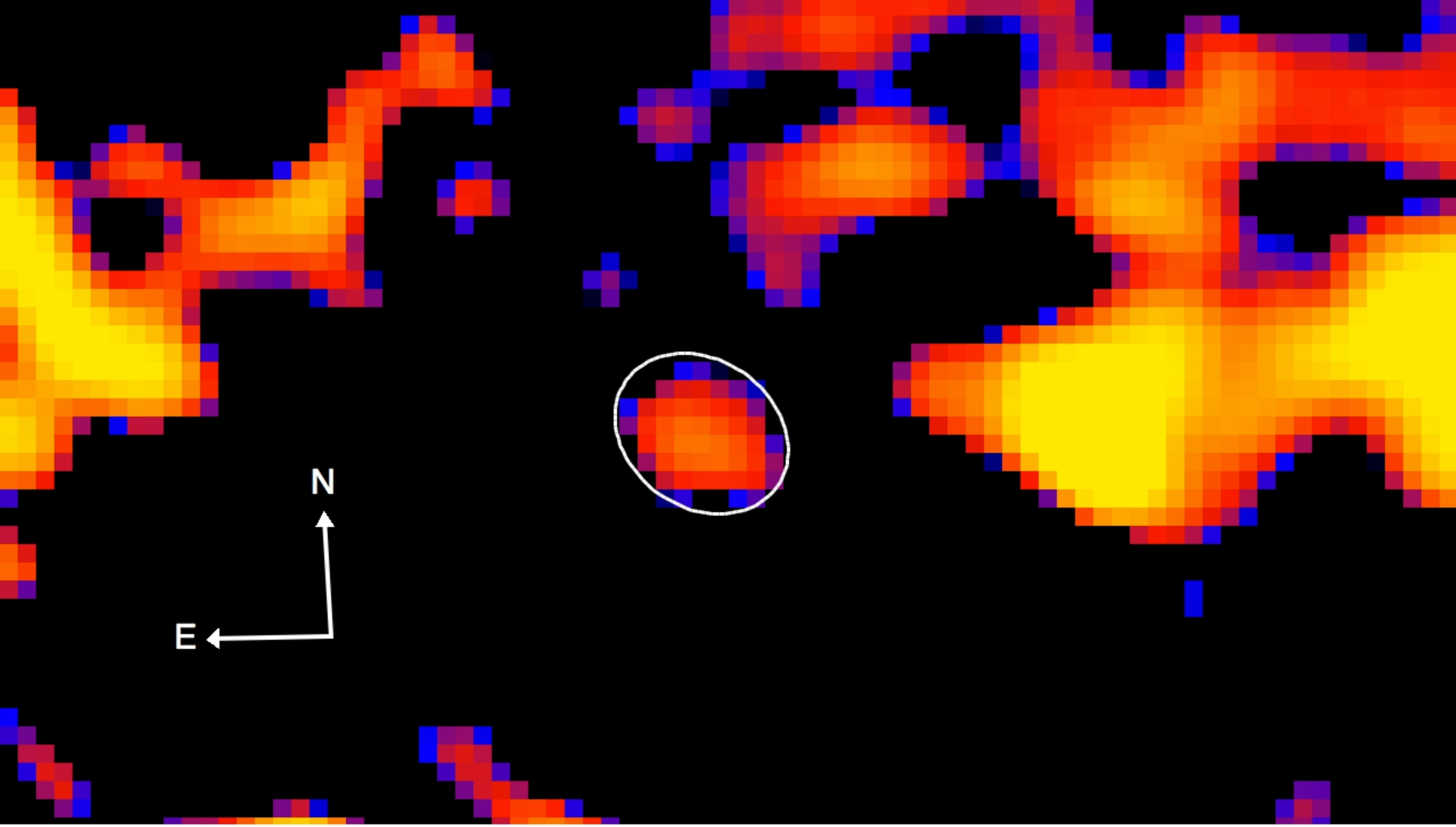}
 \caption[MGPS-2]{Image $(15\arcmin\times 11\arcmin)$ from the second epoch Molonglo Galactic Plane Survey (MGPS-2) at 843 MHz. Colors are in square root scale between $-6$ (black) and 18 (yellow) mJy/beam. The white ellipse shows the region used to extract the XMM-Newton detected diffuse source.}
 \label{fig:mgps2}
 \end{center}
\end{figure}

\section{Radio Counterpart}

The MGPS-2 is a high-resolution and large-scale survey of the galactic plane carried with the Molonglo Observatory Synthesis Telescope (MOST) at a frequency of 843 MHz  \citep{2007MNRAS.382..382M}. Close to the position of the extended X-ray source, the root-mean-square sensitivity is 1-2 mJy/beam and the beam size has a $1\sigma$ semi-axis of $20\arcsec$ (obtained by fitting a nearby point-like source).

Figure~\ref{fig:mgps2} displays the MGPS-2 image of the region around \hessj. The extended XMM-Newton source lies in a large negative area, caused by artifacts of the image reconstruction. An extended diffuse radio source, with a size corresponding very well to the XMM-Newton diffuse counterpart (the white ellipse in Fig. \ref{fig:mgps2}) is clearly observed. By comparing this source with the surrounding region, a positive excess of the order of $\sim16$ mJy/beam, can be extracted, whereas the distribution of the counts in the 4\degree$\times$ 4\degree ~image has a $1 \sigma$ width smaller than 3 mJy/beam. This radio excess, with a $1\sigma$ semi-axes of  $35\arcsec \times 26\arcsec$, corresponds well to the position and size of the extended X-ray source. Integrating this excess yields a total flux density of the order of 19 mJy. A more conservative estimate of the flux density, based on the detection threshold of the MGPS-2 survey yields to an upper limit of 25 mJy.

We also searched the Parkes 2.4 GHz survey of the southern galactic plane \citep{1995MNRAS.277...36D} performed with the Parkes radio telescope for a possible counterpart. As the resolution of the Parkes image is only 10.4\arcmin\ and the root-mean-square noise is approximately 12~mJy/beam, we could only extract an upper limit of $ \sim 100$ mJy at 2.4 GHz for the extension of the diffuse X-ray source.

\section{Searching the Infrared band using Spitzer}

Moving towards the infrared band we analyzed the GLIMPSE and MIPSGAL surveys from the Spitzer Space Telescope.
\subsection{GLIMPSE}
The Galactic Legacy Infrared Mid-Plane Survey Extraordinaire \citep[GLIMPSE,][]{2003PASP..115..953B} was performed with the IRAC instrument \citep{2004ApJS..154...10F} onboard Spitzer, which can simultaneously measure in four wavelengths at 3.6, 4.5, 5.8, and 8 $ \mu$m. The final mosaic images have a resolution of $ \sim1.2'' $. The $ 5\sigma $ point source sensitivities are 0.2, 0.2, 0.4 and 0.4 mJy for the four different wavelengths, respectively. 

We did not find any sign for diffuse infrared emission corresponding to the X-ray and radio counterparts. To determine upper limits we analyzed the infrared pixel flux distribution within and outside of the X-ray diffuse region, focusing on the 3.6 and 4.5 $ \mu$m images.

We selected all GLIMPSE pixels inside the X-ray source ellipse and subtracted the contribution of the brightest point-like sources. The final distribution was fitted with a gaussian and compared with these obtained from four other similarly shaped ellipse regions extracted in the neighborhood. The average background flux obtained from the flux distributions varies among the various ellipses indicating a gradient of infrared emission in the region.
The uncertainty on the average background flux determination ($ \leq 0.04$ MJy/sr) integrated over the ellipse area represents the minimum flux density necessary to distinguish an additional diffuse emission with respect to the other regions. For a $ 3\sigma $ detection limit, our upper limit becomes $ \sim 25$ mJy.

\subsection{MIPSGAL}

\begin{figure}[t]
 \includegraphics[width=\linewidth]{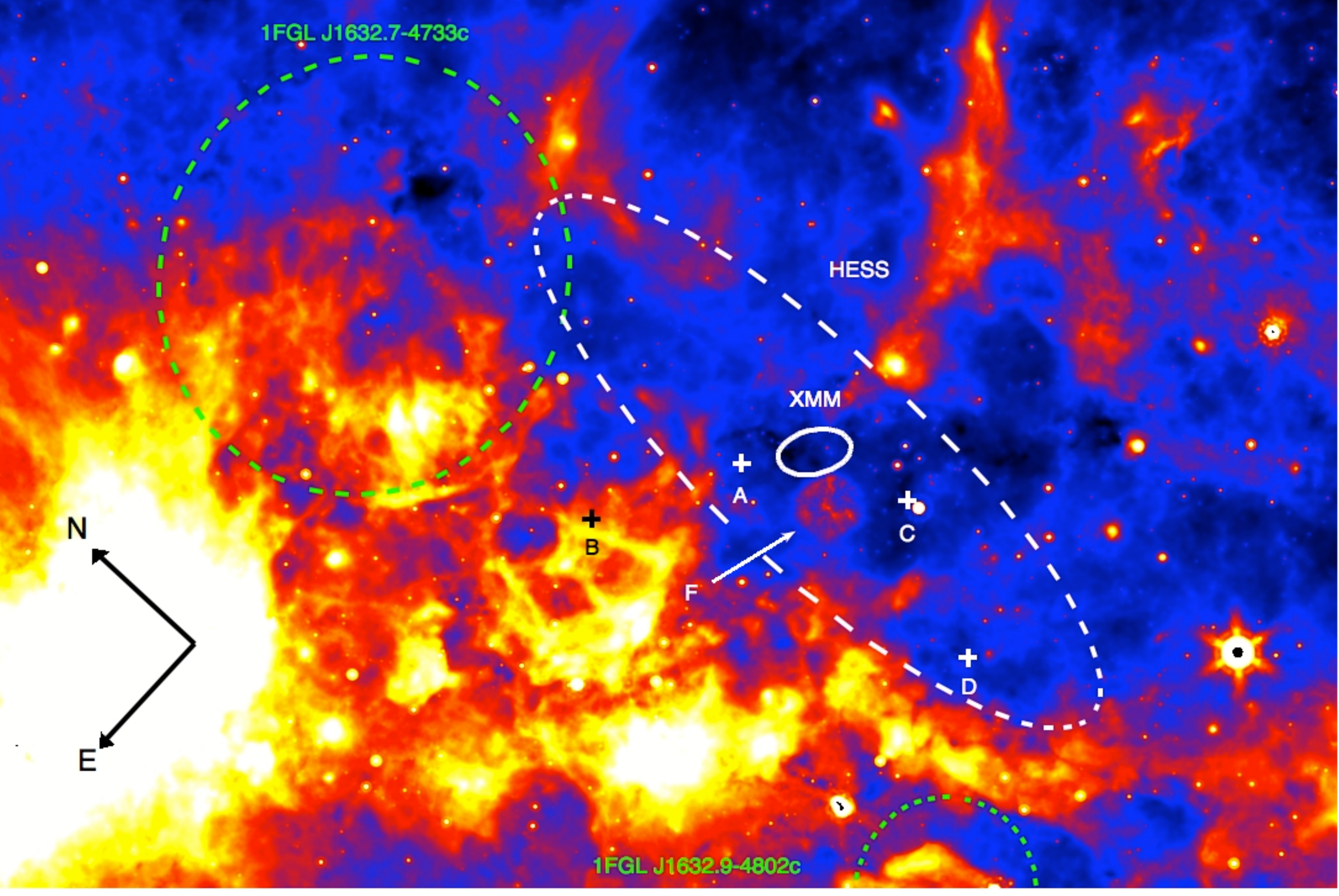}
 \caption[MIPSGAL at $ 24~\mu m $]{An extraction from the MIPSGAL survey at 24 $ \mu$m $(45\arcmin\times 30\arcmin)$. The small ellipse shows the position of the diffuse X-ray source, the dashed ellipse shows the HESS source extension \citep[as in Fig. 11 of][]{2006ApJ...636..777A} and the two large dashed circles indicates the $ 95\% $ confidence region for the FERMI detections. The crosses indicate  (A) \xmmu, (B) AXJ1632.8-4746, (C) IGR J16320-4751, (D) the nearby radio pulsar PSR J1632-4757 and (F) the unidentified 24$\mu$m circular structure.}
 \label{fig:all_reg}
\end{figure}

MIPSGAL \citep{Carey-MIPSGAL} is a survey of the inner Galactic Plane using the Multiband Infrared Photometer for Spitzer \citep[MIPS,][]{2004ApJS..154...25R}. The survey field was imaged in two passbands, 24 and 70  $\mu$m  with a resolution of $ 6'' $ and $ 18'' $, and an estimated point source sensitivities $ (3\sigma) $ of 2 and 75 mJy, respectively.  As for the GLIMPSE survey, there is no evidence for a diffuse emission corresponding to the diffuse X-ray source. In addition, at this wavelength, the infrared background is strongly inhomogeneous and does not allow to extract a useful upper limit. 

Figure~\ref{fig:all_reg} displays the $ 24~\mu$m MIPSGAL mosaic with the various point sources, diffuse X-ray source and error circles mentioned in the text. It is worth noting that an unidentified circular structure (pointed by the ``F" arrow), with a radius of $ \sim 66$\arcsec, visible at $ 24~\mu$m and not in the other bands, lies less than 93\arcsec\ away from the diffuse X-ray source. This object was detected by \cite{2009AAS...21431604F} but they have not found any counterpart. The nature of this object is unclear but its ``monochromatic" spectrum does not favor non thermal emission and a Supernova remnant origin. In the 70 and 160 $\mu$m  there is no evidence for any emission coming from the extended X-ray source nor from the 24 $\mu$m circular structure (source ``F'').

It is also interesting to remark that the extension of the TeV source lies in a region 
of faint emissivity at 24 $\mu$m (Fig. \ref{fig:all_reg}). Perhaps this infrared cavity/edge corresponds to dust blown/heated by the 
progenitor supernova or high-energy photons.

\section{The GeV sources in the vicinity}

We looked for possible  high energy counterpart  in the Fermi-LAT first year Catalogue \citep{2010ApJS..188..405A}, and  found two unidentified GeV sources in the neighborhood of \hessj\ (see Fig.~\ref{fig:all_reg}). Both these sources are flagged with the letter ``\texttt{c}'', indicating that they are to be considered as potentially confused with interstellar diffuse emission or perhaps spurious detection. Their location, flux and spectrum may not be reliable. \object{1FGL J1632.9-4802c} and \object{1FGL J1632.7-4733c} lie $15.5\arcmin$ and respectively $16.3\arcmin$ far from the extended source detected by XMM-Newton. As reported in \cite{2010ApJS..188..405A}, the diffuse background model in this region of the galaxy needs to be improved and the position of \object{1FGL J1632.7-4733c} varies with that model. It is therefore very unclear if any of these Fermi sources could be a counterpart of \hessj. The HESS error ellipse is parallel to the line joining the two Fermi sources, lying close to the extremity of that ellipse. If the two Fermi sources are real and radiate in the TeVs, the extension and inclination of \hessj\ could be significantly affected by confusion. 

\section{Discussion}

\subsection{Spectral Energy Distribution}

The match in position and size of the radio excess and of the extended X-ray source, suggests a non thermal synchrotron source emitting from the radio to the X-rays.
The TeV centroid of the H.E.S.S. source lies within the extended source detected by XMM-Newton. All other X-ray sources are further away and point-like.
This positional match and the fact that the TeV source is several times larger than the X-ray source, suggest that the non thermal synchrotron source also emits the VHE emission through inverse Compton processes.

\begin{figure}[t]
 \begin{center}
 \includegraphics[width=0.5\textwidth]{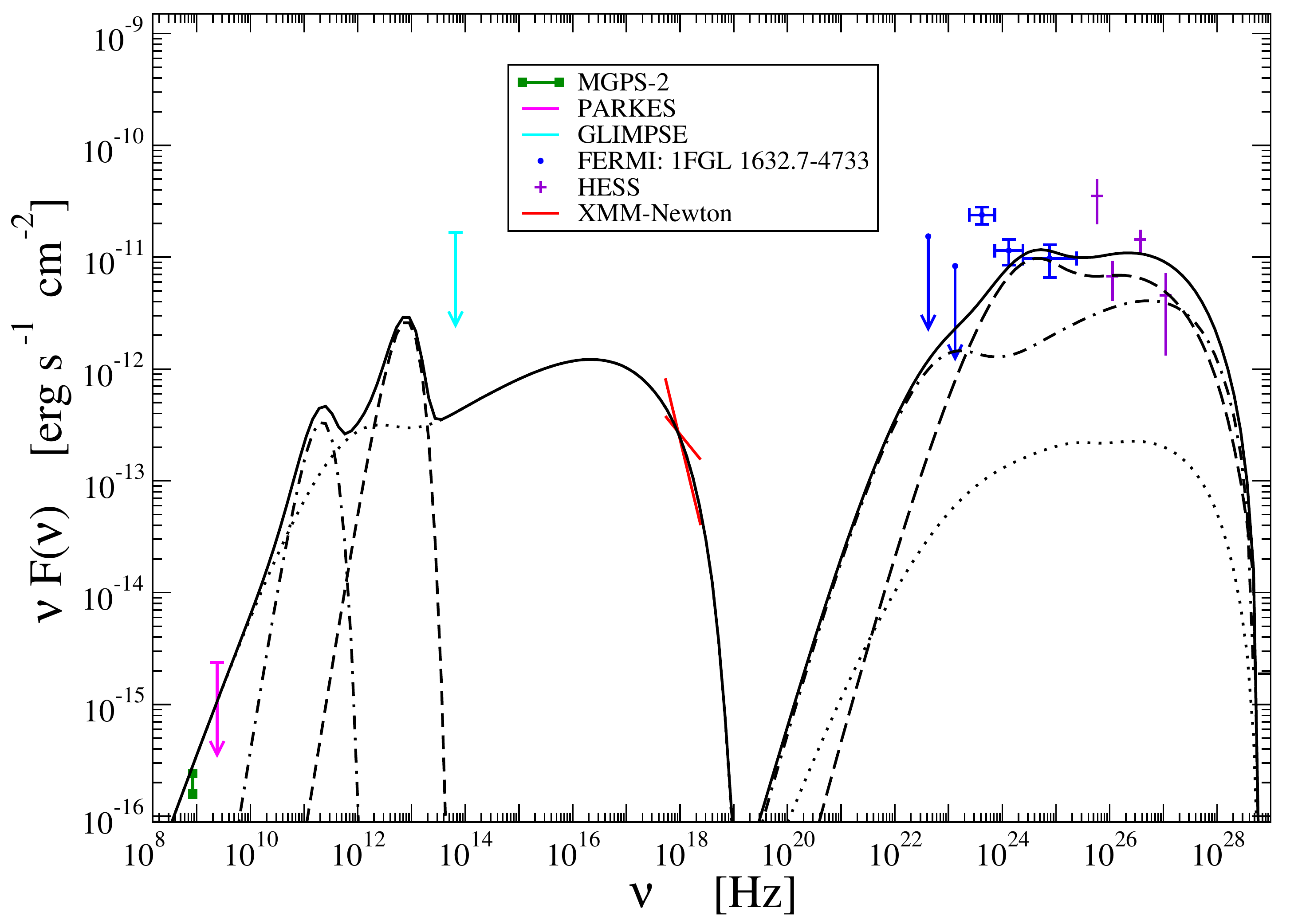}
 \caption[sed]{Spectral energy distribution of HESS J1632-478, including the upper limits from PARKES (magenta) and GLIMPSE (cyan), the probable detection from MGPS-2(green), and the detections by XMM-Newton (red) and HESS (purple). The Fermi spectrum of the nearby source \object{1FGL 1632.7-4733} is also shown (in blue). The continuous line indicates the prediction of the model used to represent the emission. At low energy the dotted line represents the synchrotron emission from the electron distribution described by Eq.~(\ref{eq:electron_distr}), the dotted-dashed and the dashed lines show the CMB and IR dust photons components respectively.
At high energy, the dotted line represents the SSC component, the dotted-dashed and the dashed lines show the external Compton emission on the CMB and dust photons components respectively.}
 \label{fig:sed}
 \end{center}
\end{figure}

The spectral energy distribution of the source, constructed with the H.E.S.S. and XMM-Newton spectra, infrared upper limits and radio detection or upper limits, is featured on Fig. \ref{fig:sed}. The significant GeV fluxes extracted from the Fermi catalogue \citep{2010ApJS..188..405A} for the close-by Fermi source FGL J1632.7-4733c are also indicated. Together, the XMM-Newton and H.E.S.S. spectra clearly indicate the presence of two spectral bumps matching the expected synchrotron and inverse Compton emission of a Pulsar Wind Nebula.

If d$_3$ is the distance of the source in unit of 3 kpc, the luminosities of the  X-ray extended and point sources are $\sim 4\times 10^{32} {\rm d_{3}^2}$ and  $2\times 10^{32}{\rm d_{3}^2}$ erg/s, respectively. The bolometric luminosity of the synchrotron and inverse Compton components are $10^{34}{\rm d_{3}^2}$ and $10^{35}{\rm d_{3}^2}$ erg/s, respectively, assuming the spectrum obtained in Fig.~\ref{fig:sed}. Using the empirical relationship of \cite{2002A&A...387..993P}, which presents a high dispersion, the X-ray luminosity is as expected for a pulsar with a total spin down power of ${\rm L_{SD}} \sim 10^{36} {\rm d_{3}^{1.5}}\ {\rm erg/s}$. The ratio of the gamma-ray and X-ray fluxes also suggest a pulsar spin-down power of $\sim3\times 10^{36}\ {\rm erg/s}$ and an age of $\sim 2\times 10^{4} \ {\rm years}$ \citep{2009ApJ...694...12M}. With 10\% of the current spin down luminosity radiated at very high energies, HESS J1632-478 is among the oldest and most gamma-ray loud known pulsar wind nebulae \citep{2007ApJ...660.1413K}, similar as these detected in \object{HESS J1825-137} \citep{2008ApJ...675..683P} or \object{HESS J1640-465} \citep{2009ApJ...706.1269L}.

Although not well defined, the TeV extension of the pulsar wind nebula is several times larger than the extension of the X-ray nebula, as observed for several aging pulsar wind nebula \citep[\object{HESS J1825-135}, \object{HESS J1420-607}, \object{HESS J1640-465};][]{2006A&A...460..365A, 2006A&A...456..245A, 2009ApJ...706.1269L}. This is usually explained by the much shorter lifetime of the electrons emitting synchrotron in the X-ray band compared to that of the electrons emitting inverse Compton in the TeVs. The very high energy synchrotron emitting electrons indeed do not have the time to reach the outer parts of the nebula. In the case of \hessj, the TeV extension is larger than the X-ray one, as observed in other sources. Note that the value of the ratio between the two components extensions can not be well constrained due to the low TeV spatial resolution and that a better determination could be achieved when more sensitive/higher resolution Cherenkov instruments will become available and possible confusion with the other high energy sources detected in the field with Fermi (see Fig. \ref{fig:all_reg}) could be resolved.

\subsection{Nature of the emission}

An upper limit to the energy density of the synchrotron photons in the nebula can be estimated, using the peak of the maximum
synchrotron emission compatible with the observed spectral energy distribution $(F_{\rm sync} < 10^{-10}\ {\rm erg\ s}^{-1}\ {\rm cm}^{-2})$ and a lower limit of the angular size of the nebula $({\footnotesize \diameter} \sim 2 \arcmin)$, as
\begin{equation}
{\rm U}_{\rm sync}=\frac{\rm L_{sync}}{\rm 4\pi c\ R^2}=\frac{\rm F_{sync}}{\rm c\ tan^2({\footnotesize \diameter}/2)}< 0.02\ {\rm eV/cm^3,}
\label{eq:U_sync}
\end{equation}

\noindent where R is the physical size of the nebula. This energy density is negligible when compared to that of the CMB (U$_{\rm CMB}$=0.25 eV/cm$^3$) or that of the infrared galactic background in the galactic plane \citep[U$_{\rm rad}\approx$ few eV/cm$^3$;][]{2006ApJ...640L.155M} which will dominate the seed photons of the Inverse Compton component.

Assuming that the same electron population is responsible for the synchrotron and for the inverse Compton emission, the ratio of the two components luminosities is given by the ratio of the magnetic and radiation energy densities
\begin{equation}
\frac{\rm L_{sync}}{\rm L_{IC}}= \frac{\rm B^2/2\mu_0}{\rm U_{rad}},
\label{eq:Ratio_synch_IC}
\end{equation}

\noindent thus the magnetic field in the nebula can be estimated as 
\begin{equation}
{\rm B}= \left(2\mu_0 {\rm U}_{\rm rad}\frac{F_{\rm sync}}{F_{\rm IC}}\right)^{0.5}= 
\sqrt{F_{\rm sync}/F_{\rm IC}}\times 10\ \mu {\rm G}.
\label{eq:B}
\end{equation}

As the spectral energy distribution qualitatively indicates L$_{\rm sync}< {\rm  L_{IC}}/10$ (considering the highest 
energy electrons, in the synchrotron nebula), the average magnetic field in the nebula is B $\sim 3\ \mu{\rm G}$. 
The spectral energy distribution also indicates that the cutoff of the synchrotron emission is in the
range $10^{16-18}$ Hz. The energy of the electrons emitting at that maximum is in the range
\begin{equation}
\gamma_{\rm max}= \left( \frac{\rm 2\pi\ m_e c\ \nu_{sync}}{\rm e\ B}\right)^{0.5}=(1-10)\times 10^{7}.
\label{eq:gamma_max}
\end{equation}

The maximum of the inverse Compton emission is  therefore 
expected at $\sim  (1-10)\times 10^{27}\ {\rm Hz}$, in agreement with the TeV spectral energy distribution. 

A possible phenomenological description of the electron energy distribution is given by the
model recently presented by \cite{Spit2008}. This author, starting from a numerical
two-dimensional particle in cell simulation, found a spectrum that, at the downstream
of the shock front, can be described by a relativistic Maxwellian plus a power-law
cut-off high energy tail. Such a model has been recently adopted by \cite{Fang2010}
to model the broad band SED of a  sample of PWNs. We adopt the same analytical model
that reads:

\begin{equation}
n(\gamma)= \left\{
\begin{array}{l}
K~\left(\frac{\gamma}{\gamma_{c}}\right)\exp\left(\frac{\gamma}{\gamma_{c}}\right)=n_0(\gamma)~~~~~~~~~ \gamma \leq \gamma_1 \\ 
n_0(\gamma)+K_1\left(\frac{\gamma}{\gamma_{c}}\right)^{-\alpha}\exp\left(\frac{\gamma}{\gamma_{c1}}\right)~~~~ \gamma >\gamma_1,
\end{array}
\right.
\label{eq:electron_distr}
\end{equation}

where $\gamma_c$  is the typical Lorentz factor of the electrons at the termination shock,
$\alpha$ is the spectral index of the high-energy tail, with a cut-off Lorentz factor $\gamma_{c1}$.
We fix $\gamma_1=7\gamma_{c}$ according to \cite{Spit2008} and $K_1$ is normalized in order to
match the Maxwellian and the high-energy tail.

We set the Lorentz factor for the particles at the termination shock to $\gamma_c=1.6\times 10^{5}$,
and use the same value of $\alpha=2.4$ as obtained in \cite{Spit2008}.
In order to match the X-ray data we use $\gamma_{c1}=1.6\times 10^{8}$.
The resulting  spectral energy distribution, reproduced using a well tested
code \citep{Tramacere2007PhD,Tramacere2009} is reported in Fig. \ref{fig:sed}
together with the observed data. We reproduce both the synchrotron self Compton (SSC) and
the external Compton emission, assuming a one-zone (post terminal shock) homogeneous 
emitting region. The details of the best-fit model parameters are reported in table \ref{tab:param}.
These parameters are only representative and a simpler model could be fitted to the data as well.
The quality and the uncertainties on the high energy data are such that deriving accurate information
on the electron distribution is not possible.

Although the Fermi data are partly compatible with the emission model, it is difficult to say if
the soft GeV spectrum of 1FGL 1632.7-4733c is related to the pulsar wind nebula or should be
considered as upper limits. Indeed, the H.E.S.S./XMM source lies well outside of the 95\% confidence 
region of 1FGL 1632.7-4733c (Fig. \ref{fig:all_reg})

With the shape of the electron distribution inferred from the external Inverse Compton component, 
the strength of the synchrotron component is fairly well constrained by the X-ray detection and by
the radio detection/upper limits.

The total energy in the electron distribution amounts to $10^{48}$ erg. This 
energy is comparable to the product of a spin-down pulsar power at birth of $10^{38}$ erg/s
and of a characteristic decay time of hundreds of years.

The nebula magnetic field, size, luminosity and inferred age are in reasonable agreement with the
simulations by \cite{Fang2010}. 

\begin{table}[h]
\caption{Emission model representing the spectral energy distribution of \hessj\ (assuming a distance of 3kpc).}
\label{tab:param}
\centerline{
\begin{tabular}{l|l|rl}
\hline \hline \noalign{\smallskip}
&Parameter & Value  &\\
\noalign{\smallskip\hrule\smallskip}
Electron distribution &$\gamma_{\rm c}$   &$ 1.6\times 10^5$  &\\
&$\gamma_{\rm c1}$  &$ 1.6\times 10^8$ &\\
&$\alpha$           &2.4&\\
&N & $7.0\times10^{-8}$ &e$^{-}$/cm$^3$\\
&R&1.6$\times 10^{18}$ &cm\\
Synchrotron&B & 1  &$\mu$G\\
External Compton&U$_{\rm rad}^{\rm CMB}$ & 0.25 &eV/cm$^3$\\
&U$_{\rm rad}^{\rm IR}$ & 2 &eV/cm$^3$\\
\noalign{\smallskip\hrule\smallskip}
\end{tabular}
}
\end{table}

\subsection{Absorption and distance}

The absorbing column densities measured by XMM-Newton in the field are much in excess to the values
N$_{H_{gal}}=(1.5-1.8)\times 10^{22}$ cm$^{-2}$ derived from the low resolution radio surveys \citep{2005A&A...440..775K,1990ARA&A..28..215D}.

The absorbing column density on the pulsar wind nebula (N$_H \approx 11\times 10^{22}$ cm$^{-2}$) is very similar to that observed for IGR~J16320-4751 and slightly larger than these
observed in the directions of AX J1632.8-4746  and \xmmu\  (N$_H \approx 6\times 10^{22}$ cm$^{-2}$). This
indicates an unusually large column density in this field and that about half of the absorption towards
IGR~J16320-4751 \citep{2006A&A...453..133W} may not be related to the High-Mass X-Ray Binary system.
It also indicates that the distance to \hessj\ is of the same order as that of IGR~J16320-4751, which was
approximately estimated as 3.5 kpc \citep{2008A&A...484..801R}.

\section{Conclusions}

We observed the unidentified TeV source \hessj\ with XMM-Newton and looked for counterparts in the
GeV, infrared and radio bands. An extended faint X-ray source is detected close to the centroid of the H.E.S.S. error ellipse. A radio excess corresponding to the X-ray source is found in the Molonglo sky survey. Upper limits have been derived from Spitzer and Parkes data. The GeV image obtained by Fermi shows two close-by sources flagged as confused in the Fermi catalogue, but none of them corresponds to the X-ray source, the situation is therefore unclear.

The flux density emitted by \hessj\ at very high energies is at least 20 times larger than observed at the other wavebands probed. The source shape and spectral energy distribution suggests a pulsar wind nebula and can be used to successfully constrain a one zone model for the post terminal shock region of the pulsar wind nebula.
The assumed relativistic electron distribution is Maxwellian $(\gamma\sim 10^5)$ with a non thermal
tail extending to  $\gamma\sim 10^8$. The synchrotron nebula is faint because of the low 
magnetic field (3 $\mu$G).

The point-like X-ray source, detected in the synchrotron nebula, is probably the signature of the pulsar and of the termination shock. The age of the pulsar is estimated as some $10^4$ years. The lack of spatial resolution and the probable confusion of the GeV/TeV sources in this field does not allow to perform a detailed study of the inverse Compton emitting region and of its interactions. More sensitivity and spatial resolution are needed at very high energies.

\begin{acknowledgements}
Based on observations obtained with XMM-Newton, an ESA science mission with instruments and 
contributions directly funded by ESA Member States and NASA. The Molonglo survey is supported by 
the Australian Research Council. J.A.Z.H. acknowledges the Swiss National Science Foundation for financial support.
\end{acknowledgements}

\bibliographystyle{aa}
\bibliography{14764}
\newpage

\mbox{}

\newpage

\end{document}